\documentclass[12pt,preprint]{aastex}
\usepackage{graphicx}
\usepackage{epsfig}
\usepackage{epstopdf}

\def \arcsec {\hbox{$^{\prime\prime}$}}

\def \hic   {{\em Hi-C}}
\def \sdo   {{\em SDO}}
\def \hinode   {{\em Hinode}}
\def \aia   {{\em AIA}}
\def \hmi   {{\em HMI}}
\def \xrt   {{\em XRT}}
\def \trace   {{\em TRACE}}
\def \nixt   {{\em NIXT}}

\def \fex     {Fe\,{\sc x}}

\def \fexviii {Fe\,{\sc xviii}}

\begin{document}

\shorttitle{Observing coronal nanoflares in active region moss}
\title{Observing coronal nanoflares in active region moss}
\author{Paola Testa\altaffilmark{1}, Bart De Pontieu\altaffilmark{2},
  Juan Mart\'inez-Sykora\altaffilmark{2,3}, Ed DeLuca\altaffilmark{1},
  Viggo Hansteen\altaffilmark{4}, Jonathan Cirtain\altaffilmark{5},
  Amy Winebarger\altaffilmark{5}, Leon Golub\altaffilmark{1}, Ken
  Kobayashi\altaffilmark{5}, Kelly Korreck\altaffilmark{1}, Sergey
  Kuzin\altaffilmark{6}, Robert Walsh\altaffilmark{7}, Craig
  DeForest\altaffilmark{8}, Alan Title\altaffilmark{2}, Mark
  Weber\altaffilmark{1}}

\affil{\altaffilmark{1} Smithsonian Astrophysical Observatory,60 Garden street, MS
  58, Cambridge, MA 02138, USA; ptesta@cfa.harvard.edu} 
\affil{\altaffilmark{2}Lockheed Martin Solar \& Astrophysics Lab, Org.\ A021S,
  Bldg.\ 252, 3251 Hanover Street Palo Alto, CA~94304 USA}
\affil{\altaffilmark{3} Bay Area Environmental Research Institute,
  Sonoma, CA 95476 USA} 
\affil{\altaffilmark{4} Institute of Theoretical Astrophysics, University of Oslo,
  P.O. Box 1029 Blindern, N-0315 Oslo, Norway} 
\affil{\altaffilmark{5} NASA Marshall Space Flight Center, ZP 13, Huntsville, AL 35812} 
\affil{\altaffilmark{6} P.N.Lebedev Physical institute of the Russian
  Academy of Sciences, Leninskii prospekt, 53, 119991, Moscow} 
\affil{\altaffilmark{7} University of Central Lancashire, Preston,
  Lancashire, United Kingdom, PR1 2HE}  
\affil{\altaffilmark{8} Southwest Research Institute, 1050 Walnut
  Street, Suite 300, Boulder, CO 80302}

\begin{abstract}
  The High-resolution Coronal Imager (\hic) has provided Fe XII
  193\AA\ images of the upper transition region moss at an
  unprecedented spatial ($\sim 0.3-0.4$\arcsec) and temporal (5.5s)
  resolution.  
  The \hic\ observations show in some moss regions variability on
  timescales down to $\sim 15$~s, significantly shorter than the minute
  scale variability typically found in previous observations of moss,
  therefore challenging the conclusion of moss being heated in a
  mostly steady manner.  These rapid variability moss regions
  are located at the footpoints of bright hot coronal loops observed
  by \sdo/\aia\ in the 94\AA\ channel, and by \hinode/\xrt.
  The configuration of these loops is highly dynamic, and suggestive 
  of slipping reconnection.
  We interpret these events as signatures of heating events associated
  with reconnection occurring in the overlying hot coronal loops,
  i.e., coronal nanoflares. 
  We estimate the order of magnitude of the energy in these events to
  be of at least a few $10^{23}$~erg, also supporting the nanoflare
  scenario.   
  These \hic\ observations suggest that future observations at
  comparable high spatial and temporal resolution, with more extensive
  temperature coverage are required to determine the exact characteristics
  of the heating mechanism(s).
\end{abstract}


\section{Introduction} 
\label{sec:intro}
Coronal heating is one of the central problems in solar physics, and
the detailed characteristics of the processes converting magnetic into
thermal energy are still hotly debated \citep[see e.g.,][for
reviews]{Klimchuk06,Reale10}.
Heating processes are likely to occur on small spatial and temporal
scales, which are difficult to access with the typical resolution of present
instrumentation. 
The spatial and temporal characteristics of heating in active regions (ARs)
have been investigated by indirectly inferring constraints from the
spatial, temporal and thermal properties of the plasma from imaging
and spectral data, and comparing them with predictions of different
heating models
\citep[e.g.,][]{Reale09b,Reale09,McTiernan09,Hansteen10,Testa11,Tripathi11,Winebarger11,Warren12}.   
One of the candidate processes for coronal heating is the nanoflare
model, where the heating is produced by reconnection due to braiding
of coronal magnetic field lines caused by random motions of the loop
footpoints in the photosphere \citep{Parker88,Cirtain13}.  Several models have
been developed assuming nanoflare events to occur in unresolved loop
strands \citep[e.g.,][]{Cargill94,Cargill97}, and provide a viable
interpretation of various coronal AR observations which are
constraining the spatial distribution and frequency of the heating events
\citep[e.g.,][]{Reale11,Warren11}.  However, past observations provide
only limited constraints to the models, and higher quality observations
are crucial to test the models.
We note that scenarios alternative to Parker's have been
  proposed, which can also produce nanoflare-like heating
 \citep[e.g.,][]{vanballegooijen11,depontieu11}.

Heating events are difficult to investigate in the corona, where
conduction is extremely efficient in smoothing out gradients,
therefore many studies have focused on ``moss'', i.e., the upper
transition region (TR) layer of high pressure loops in active regions,
which is very bright in narrowband EUV images sensitive to $\sim 1$~MK
emission \citep[e.g.,][]{Peres94,Fletcher99,Berger99,
  depontieu99,Martens00}.  
Revealed first by recent high spatial resolution X-rays and EUV
narrowband coronal imagers, such as \nixt\ \citep{nixt} and \trace\
\citep{trace}, the moss is very bright in cool EUV bands (e.g.,
171\AA\ and 193\AA). Moss is generally relatively free from 
line-of-sight (LOS) confusion due to the lack of overlying
coronal emission, since the overlying coronal loops are hotter (few
MK) and only weakly emitting in cool EUV narrowbands. 
The moss has been found to have typical temperatures of
0.6-1.5~MK, densities of the order of $10^{10}$~cm$^{-3}$, and
  thickness of about 1-3~Mm
  \citep[e.g.,][]{Fletcher99,Berger99,Warren08}.  
Studies using imaging and spectral observations of moss have found
relatively low level temporal variability, placing constraints on the
frequency and distribution of intensity of heating events, and
it has often been interpreted as an indication of quasi-steady heating
of the moss and therefore of the associated AR core loops
\citep{Antiochos03,Brooks09,Tripathi10}. However, spatial and temporal
averaging can significantly influence the results found for moss
variability, as we show here.  

In this Letter, we investigate the properties of moss by using EUV
imaging data of the High-resolution Coronal Imager (\hic) instrument,
which are characterized by unprecedented spatial ($\sim 0.3-0.4$\arcsec)
and temporal ($\sim 5.5$~s) resolution. 
\hic, launched on a sounding rocket on 2012 July 11, has provided about
5 minutes of EUV observations of coronal and TR plasma
in a 6.8 arcmin $\times$ 6.8 arcmin field of view, in a narrow
passband very similar to the \sdo/\aia\ 193\AA\ band
\citep{Cirtain13}. 
We show that the spatial and temporal resolution of \hic\ data reveal
moss variability on very short timescales, contrary to previous
findings, and suggesting a more episodic nature of the heating
compared to what was previously thought. These rapid and intense
brightenings are found at the footpoints of hot (several MK) loops,
and we interpret them as the signature of heating episodes due to
reconnection higher up in the corona, i.e., coronal nanoflares.  
In order to investigate the relation of the high variability in
the moss with the overlying coronal emission and surface magnetic
field we analyze simultaneous observations from \hic, \sdo/\aia\
\citep{Lemen12}, \hinode/\xrt\ \citep{Golub07} and \sdo/\hmi\
\citep{hmi}, as described in Section~\ref{sec:observations}.  
In Section~\ref{sec:results} we present the results of our analysis of
the temporal variability in the moss as observed by \hic, and spatial
correlations with hotter corona emission and photospheric magnetic
field.  We discuss our findings and draw our conclusions in
Section~\ref{sec:discussion}.

\section{Observations}
\label{sec:observations}

\begin{figure*}[!ht]
\centerline{\includegraphics[scale=0.6]{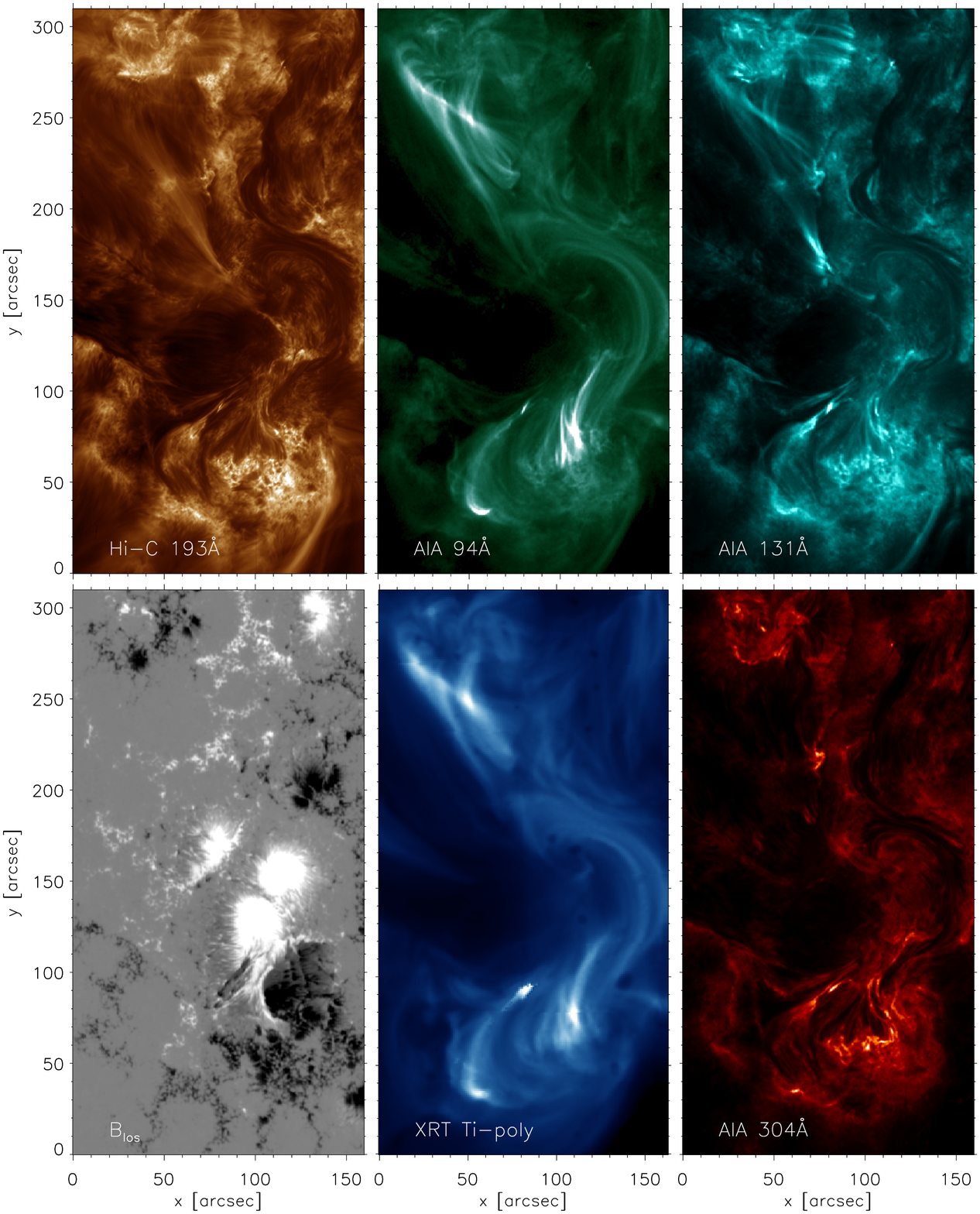}}\vspace{-0.9cm}
\caption{\hic, \sdo\ and \hinode/\xrt\ observations of AR 11520 on
  2012 July 11. We show images obtained by averaging
  intensities in the interval approximately between 18:52:50UT and
  18:55:30UT. Solar north is to the right, and solar east is down.
  Top row, left to right: \hic\ 193\AA\ passband,
 \sdo/\aia\ 94\AA\ and 131\AA\ passbands. Bottom row, left to right:
 \sdo/\hmi\ magnetic field, \hinode/\xrt\ Ti-poly, \sdo/\aia\
 304\AA.\label{fig:hic_aia}} 
\end{figure*}

We analyze \hic\ observations ($\sim 0.103$\arcsec\ pixel$^{-1}$) of
AR 11520 on 2012 July 11 around 18:53UT. 
Figure~\ref{fig:hic_aia} shows the region we selected ($\sim
160$\arcsec $\times 310$\arcsec) from the full field of view, to focus
on the properties of moss regions.  
We also analyze simultaneous imaging data of the same region in
\sdo/\aia\ EUV narrowbands ($\sim 0.6$\arcsec\ pixel$^{-1}$, 12~s
cadence), and in \hinode/\xrt\ Ti-poly filter ($\sim 1$\arcsec\
pixel$^{-1}$, 15~s cadence).    
The \aia\ and \xrt\ data have been processed with the aia\_prep
and xrt\_prep routines respectively, available as a part of SolarSoft.

The \aia, \xrt, and \hmi\ data have been rescaled to the \hic\ spatial
resolution, corrected for solar rotation, and rotated and
transposed to match the \hic\ data.  A time series in each band,
matching the higher \hic\ temporal resolution, has been created by
taking the corresponding image closest in time to the \hic\ observing
times.  
For \xrt\ we combined short and long exposure images to obtain the
best signal-to-noise ratio in both dark and bright regions, while avoiding
saturation. 
 The data from different instruments have been carefully coaligned
by applying cross-correlation procedures.  
\aia\ and \xrt\ data, averaged over the duration of the \hic\
observations, as well as an \hmi\ data are shown together with \hic\
data in Figure~\ref{fig:hic_aia}. 

The comparison of the images taken by the different instruments shows
several moss regions, which are the brightest features in the \hic\
193\AA\ and \aia\ 304\AA\ and 131\AA\ images, while in \xrt\ and \aia\
94\AA\ coronal loops tend to be the most prominent features. Though 
the \aia\ 94\AA\ channel is sensitive to both cool ($\sim 1$~MK) and 
hot ($\sim 6$~MK) emission \citep[e.g.,][]{Boerner12,Testa12}, for these
loops the \aia\ 94\AA\ emission is dominated by \fexviii\ emission
as supported by the comparison of the \aia\ coronal channels
\citep[according to the method of][]{Testa12b}, their temporal
evolution, and the close morphological correspondence with the \xrt\
loops.

\section{Results} 
\label{sec:results}

The high spatial and temporal resolution observations of \hic\ show a
variety of spatial and temporal scales. \cite{Cirtain13} 
investigated brightening in twisted loops (located around $[x,y] \sim
[70, 80]$\arcsec\ in Figure~\ref{fig:hic_aia})
interpreted as a result of energy release due to fine-scale magnetic
field braiding.
Close inspection of the \hic\ movies (Movie~1 and 2) shows that a
  subset of moss regions are characterized by striking temporal
variability dominated by localized short duration brightenings ($\sim
15-20$~s).  These timescales are significantly shorter than for the
typical moss variability that previous studies found on timescales
of the order of minutes, and interpreted as due to absorption of
EUV emission by chromospheric dynamic fibrils
\citep{depontieu03b,depontieu03}. 

\begin{figure*}[!ht]
\hspace{1.2cm}\centerline{\includegraphics[scale=1]{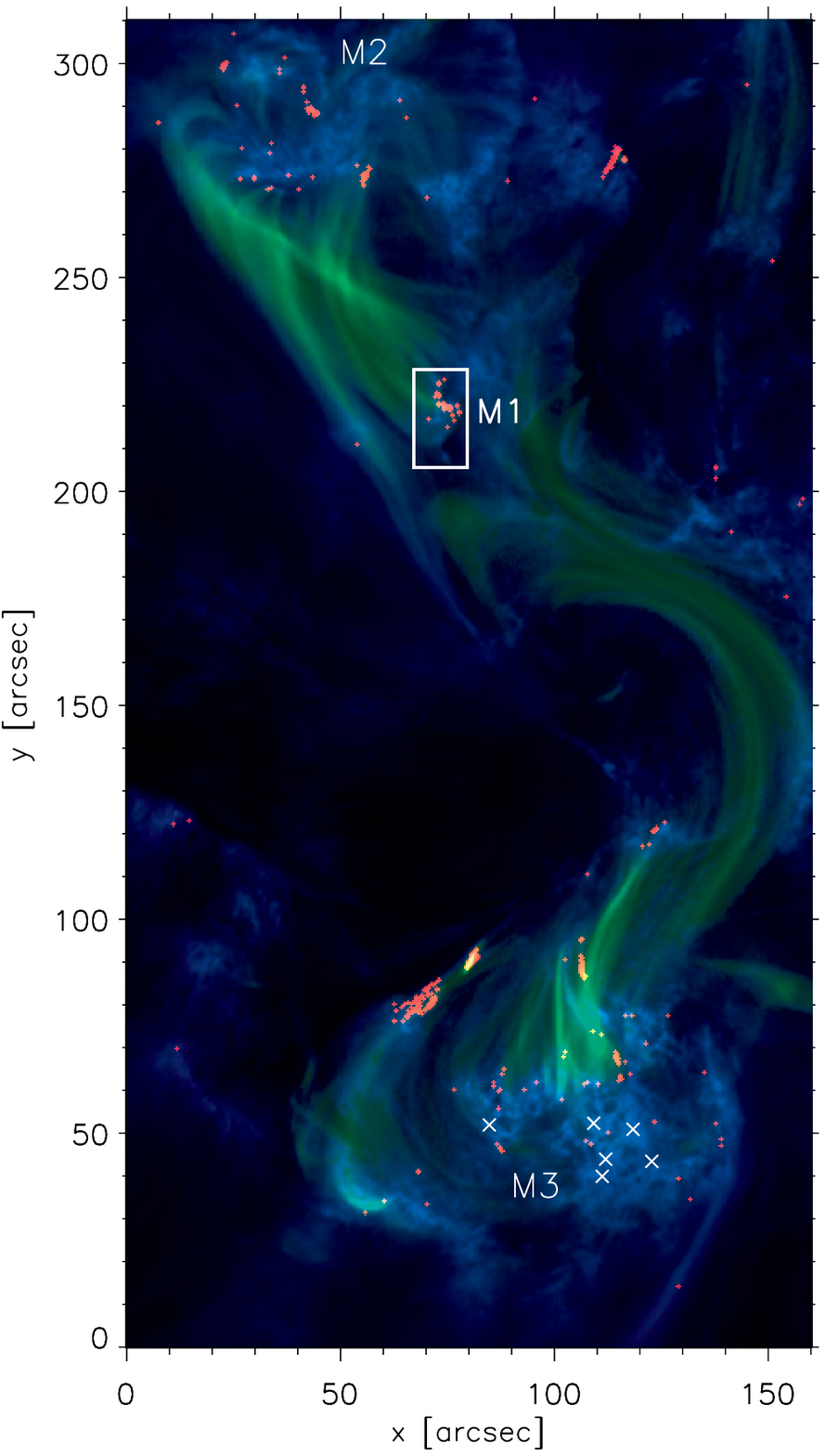}
 \hspace{-0.7cm}\includegraphics[scale=1]{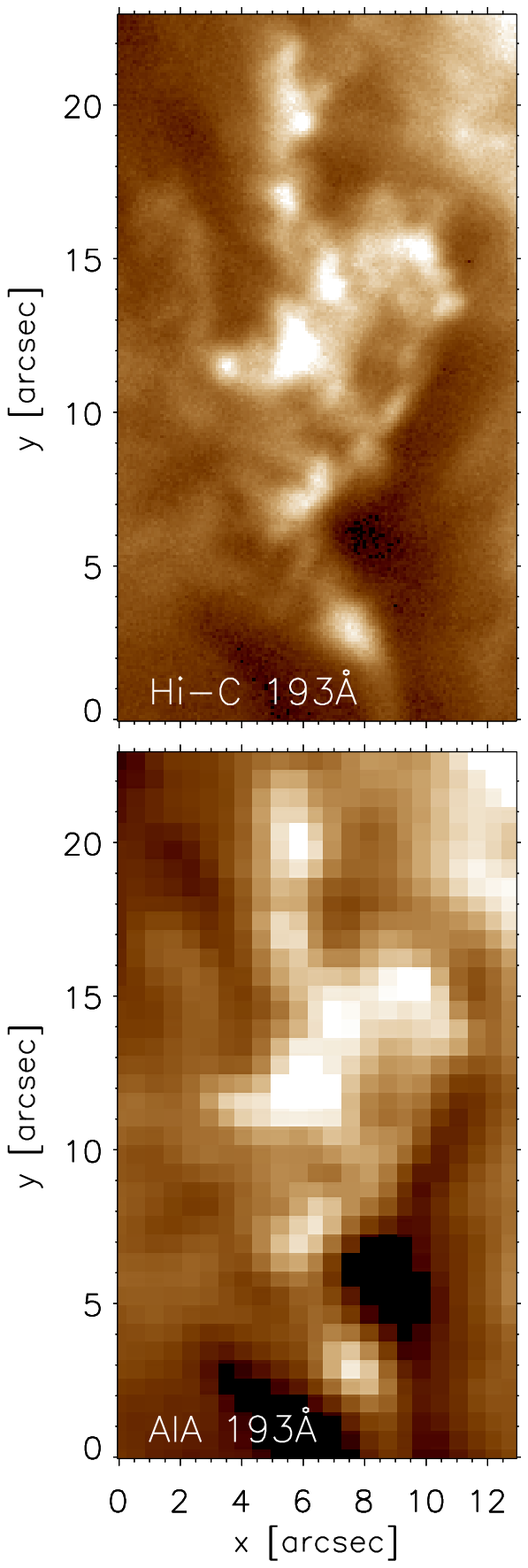}}\vspace{-0.8cm}
\caption{Left: 3 color image combining \hic\ data (blue), 
 \sdo/\aia\ 94\AA\ data (green), and a mask highlighting 
  regions with high level of variability in the \hic\ time series (red;
  see text for details). The white crosses mark moss locations outside
  high variability areas which we select for comparison with the rapidly
  varying moss (Figure~\ref{fig:hic_moss_lc}).
  Right: 193\AA\ \hic\ (top; $\sim 0.1$\arcsec\ pixel$^{-1}$) and
  \aia\ (bottom; $\sim 0.6$\arcsec\ pixel$^{-1}$) images of the rapid
  variability moss region M1 (white box in left panel), clearly
  showing the significantly different spatial resolution of the two
  instruments.\label{fig:hic_3col}} 
\end{figure*}

We applied an analysis to the \hic\ time series to single out these
regions where brightenings on short timescales occur, characterize
the properties of these regions, and investigate the reasons for their
strikingly enhanced temporal variability. We spatially
rebinned the \hic\ data to the level of the effective spatial
resolution, and derived time series of the running difference of the
rebinned data.  We then apply an algorithm calculating the number of
zero crossings of the running difference lightcurves. Finally we
obtain a mask of the points where the number of zero crossing is above
a certain threshold, since we want to highlight regions characterized
by episodic brightenings, i.e., where the lightcurve increases and then
decreases to roughly pre-brightening levels, as opposed to steady
increases on long timescales. We also impose that the level of
variability in intensity is above a given threshold 
above the average intensity. 
We explored a small range of spatial rebinning scales, temporal scales
for the running difference, and significance of variability. 
Figure~\ref{fig:hic_3col} (left panel) shows the map resulting
from our temporal analysis (red), calculated using a factor 4
spatial rebin, a running difference calculated over 3 time steps
($\sim 16.5$s), and significance above $4 \sigma$. The map is
superimposed on the \hic\ data (blue) and the \aia\ 94\AA\ data
(green). 

This map confirms the impression from the visual inspection of the
time series, showing a high concentration of locations characterized
by high variability on small time scales in the moss region M1 (white
box in Figure~\ref{fig:hic_3col}). 
We note that there appears to be a close correspondence of the rapid
variability regions with the brightest areas in the \aia\ 304\AA\
image. 
The map of high variability regions also includes a few isolated features
in both large moss regions in the field of view (labeled M2, and M3),
as well as the reconnection event studied by \cite{Cirtain13} (at $[x,y]
\sim [70,80]$), and some other coherent events like apparent flows
(e.g., around $[x,y] \sim [110,90]$ and $[x,y] \sim [115,280]$; see
Movie~1).   
In the following we focus on the very prominent short scale
variability in M1.

\begin{figure*}[!ht]
\centerline{\includegraphics[scale=0.75]{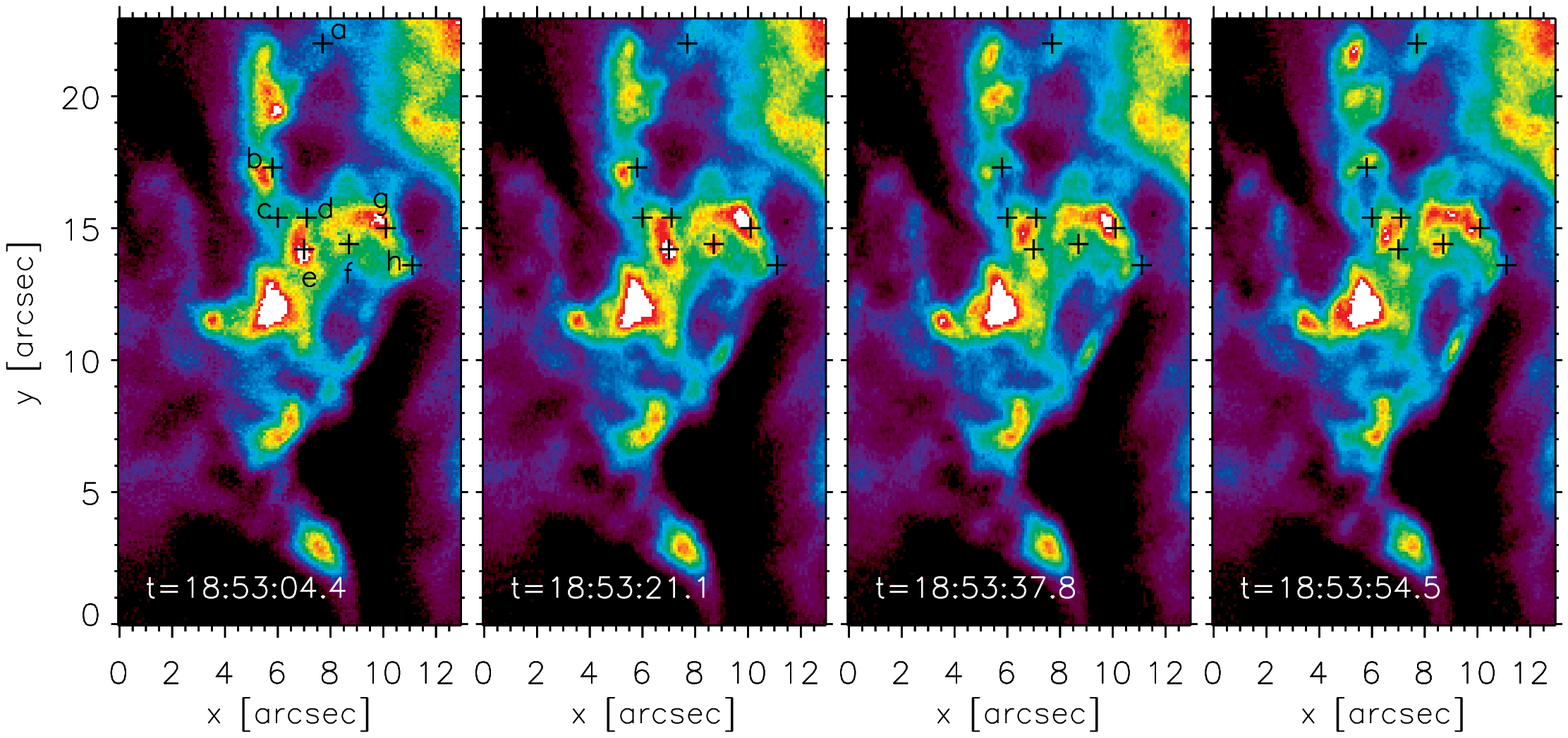}}\vspace{-0.8cm}
\caption{\hic\ images of M1, at four different times, at about
  16.5s interval from one another, showing large variations on time
  scales of about 15-30s in some locations. We label a few selected
  locations for which we show the \hic\ lightcurves in
  Figure~\ref{fig:hic_moss_lc}.\label{fig:hic_moss_var}} 
\end{figure*}

\begin{figure*}[!ht]
\centerline{\includegraphics[scale=0.78]{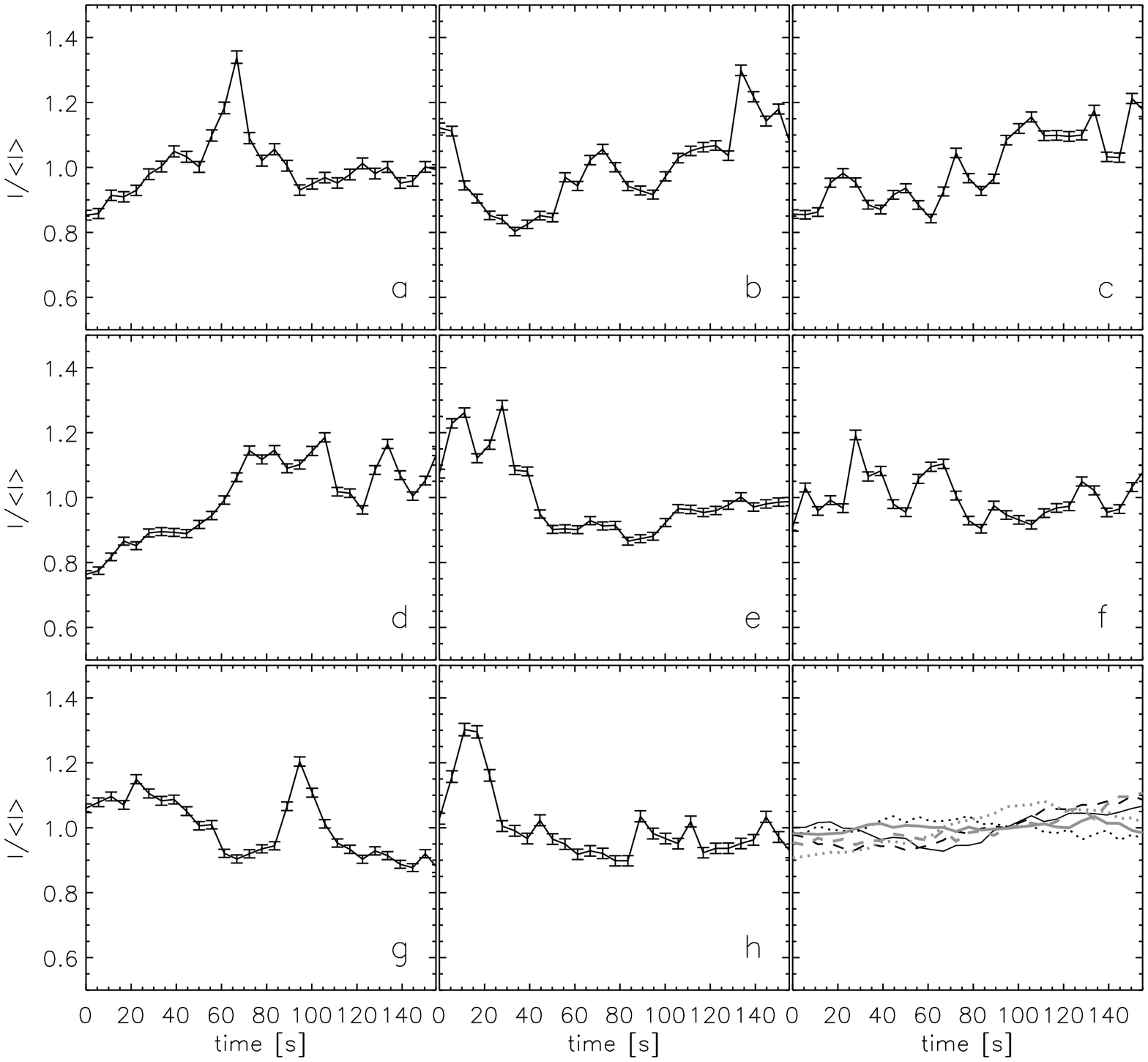}}\vspace{-1.2cm}
\caption{\hic\ normalized lightcurves for 8 locations 
  labeled ($a-h$) in Figure~\ref{fig:hic_moss_var}, and for 6 moss
  locations marked in Figure~\ref{fig:hic_3col} not characterized by
  rapid variability (bottom right panel). The lightcurves are obtained
  by integrating the intensity in a 3 pixel $\times$ 3 pixel area
  (i.e., comparable with effective spatial resolution of the instrument).\label{fig:hic_moss_lc}} 
\end{figure*}

The comparison of \hic\ and \aia\ 193\AA\ images of the moss region M1
(Figure~\ref{fig:hic_3col}, right panels), clearly illustrates the
significant improvement of \hic\ in spatial resolution.   
In Figure~\ref{fig:hic_moss_var} we show four images of the high
variability moss region M1 at about 16.5s cadence. 
The high variability is clearly evident already with this limited
temporal sampling. 
A few locations, showing rapid temporal variability, as indicated by
the temporal analysis described above, are marked by crosses and
labeled, and their lightcurves are shown in
Figure~\ref{fig:hic_moss_lc}. For comparison, we also show sample
lightcurves for 6 moss locations not characterized by rapid
variability (marked in Figure~\ref{fig:hic_3col}).

As evident from Figure~\ref{fig:hic_aia} and \ref{fig:hic_3col}, the
rapidly variable moss region is located at the footpoints of the
hottest coronal loops in the AR, which are bright in the \aia\ 94\AA\
and \xrt\ images. Inspection of the longer time series of \aia\ images
(see Movie~3) indicates that these loops become brighter relatively
shortly before the start of \hic\ observations ($\sim 20$min) and the
whole coronal loop configuration is highly dynamic. 
Sets of loops connecting the moss region M2 with M1 appear to be
crossing in the plane of the sky. The area where the hot loops appear
to cross (around $[x,y] \sim [50,250]$) shows an increase in
brightness that is followed by increased brightness along the entire
length of loops.  
The time series of the magnetic field from \hmi\ and of the coronal
emission in \aia\ bands (Movie~3 and 4)  support a scenario in
  which the pre-existing configuration of coronal loops between M1 and
  the conjugate footpoints, M2, are impacted by emerging flux, as
  evidenced by the brightening of the loops connecting the right-end
  area of M2 to this emerging flux region.  
  These hot loops are rapidly evolving, and the apparent displacement
  of their footpoints towards M1 supports a slipping reconnection
  scenario as in \cite{Aulanier07}. This
reconnection process might explain the brightening of all
hot loops anchored to the moss region M2, which is observed
simultaneously with the apparent footpoint motion. 
We discuss in the following the possible implications of these
findings.

\section{Discussion and Conclusions}
\label{sec:discussion}

In this Letter we have presented observations of active region moss
with the \hic\ narrowband EUV imager. The unprecedented combination of
high spatial and temporal resolution of \hic\ shows that the moss is
structured down to small spatial scales, like the chromosphere
\citep{depontieu03}, and it reveals high levels of temporal
variability. This challenges previous findings of moss steady emission
\citep{Antiochos03,Brooks09}. 
\hic\ shows moss areas characterized by slow variability, on typical
timescales of minutes, analogous to previous findings
\citep{depontieu03b,depontieu03}, but it also shows moss locations
with high intensity brightenings on very short timescales (down to
$\sim 15$~s).   

We investigated the characteristics of these rapid variability regions
which might explain their distinctively higher temporal variability
compared to other moss.
We found that these rapid variability regions show an intriguing
correlation with the brightest 304\AA\ emission, and they occur at
the footpoints of the hottest loops ($\sim 6$~MK), which are
brightening after flux emergence is observed. 
As shown in section~\ref{sec:results}, the evolution of the coronal
configuration of these hot loops is suggestive of slipping
reconnection.   
 Even if we cannot exclude heating in the TR region,
these observed correlations with the magnetic and coronal features
suggest that the rapid moss variability is due to reconnection events
and resulting energy release occurring in the corona, i.e., to coronal
nanoflares. The energy transport to the TR layers can be due to
either thermal conduction and/or to beams of non-thermal particles
accelerated at the reconnection site
\citep[e.g.,][]{Brosius12,BrosiusHolman12}.
We note that the signal-to-noise ratio and temporal resolution of the
\aia\ 94\AA\ images are too low to resolve these heating events in the
corona (as expected on the basis of simulations of nanoflare
  heated multi-stranded loops; e.g., \citealt{Reale08,Bradshaw11})
and to investigate their temporal correlations with the events
\hic\ observes at lower temperatures at the loop footpoints. 

 Since the TR emission timescales depends on the timescale of
  evolution of pressure \citep[see e.g.,][]{Klimchuk08}, the extremely
  short timescales for the moss brightenings imply that the associated
  heating events must last less than $\sim 15$s (comparable to
  processes observed in flares, such as tearing mode instability;
  \citealt{Kliem95,Cassak09}), providing tight constraints on heating
  theories.
The observed lightcurves imply a very short cooling time for the
emitting plasma, since this is TR emission: heating of the plasma out
of the passband is incompatible with these observations as it would
imply that the $\sim 1$~MK TR would be pushed down to even larger
densities, therefore causing a further increase in the 193\AA\
intensity.
The conductive cooling times for the loop can be written as
  $\tau_c \sim 4.6 \cdot 10^{-10} T^{-5/2} n_e L^2$
  \citep[e.g.,][]{Reale10}. Estimating for the loop semilength $L \sim
  3 \cdot 10^9$~cm, and assuming for the coronal plasma density $n_e
  \sim 10^9$, and temperature $T \sim 8$~MK, we find timescales of the
  order of 20s, comparable with the observed timescales.  

By using the constraints from the \hic\ lightcurves, we can also
estimate an order of magnitude for the energy of these events, $E =
3n_ek_BTV$, where $V = A \cdot \ell$, $A$ can be written as the
area of the \hic\ pixel times the number of pixels $n_{pix}$ involved
in the brightening event, and $\ell$ is the depth of the emitting
plasma along the LOS.
The TR density can be written as a function of the \hic\ intensity of
the brightenings $\Delta I$, since $\Delta I = \Delta EM_{los} \cdot
R_{HiC}(T)$, where $R_{HiC}(T)$ is the \hic\ temperature response
(obtained from the routine hic\_get\_response in SolarSoft), and
$EM_{los}$ is the emission measure per unit area $= n_e^2 \ell$.
Therefore, $E \propto T n_{pix} \cdot (\Delta I \cdot \ell
/R_{HiC}(T))^{1/2}$.  For $n_{pix} =9$, i.e., about the effective
spatial resolution (though some events possibly involve larger areas,
up to $\sim 5 \times 5$~pixel), $\ell = 1-3 \cdot 10^8$~cm
(see sec.~\ref{sec:intro}), we find that the energies of these events
have typical values of a few $10^{23}$~erg.   
As we are only estimating the energy deposited in the TR, this is
likely a lower limit to the total energy in these heating events.
The total energy radiated by the plasma in these events is $n_e^{2} V
\cdot P(T)  \cdot t = \Delta I  \cdot P(T)  \cdot A  \cdot t / R_{HiC}(T)$,
where $P(T)$ represents the radiative losses of a plasma at
temperature $T$ (from CHIANTI; \citealt{chianti7}), and $t$ is the
duration of the events.  We find that in the range of $\log T \sim
[5.7-6.2]$ the radiated energy is $E_r \lesssim 10^{22}$~erg,
confirming that the plasma is likely cooling more by
conduction than radiation.

In summary the data are compatible with a heating event and subsequent
conductive cooling, and the energies are of magnitude comparable with
small nanoflare events. We note that all of these estimates can be
affected by other factors such as absorption of EUV emission by
chromospheric material \citep{depontieu09b}, or the element abundances
of the emitting plasma \citep[e.g.,][]{Testa10ssrv}. 

\begin{figure}[!ht]
\centerline{\includegraphics[scale=0.55]{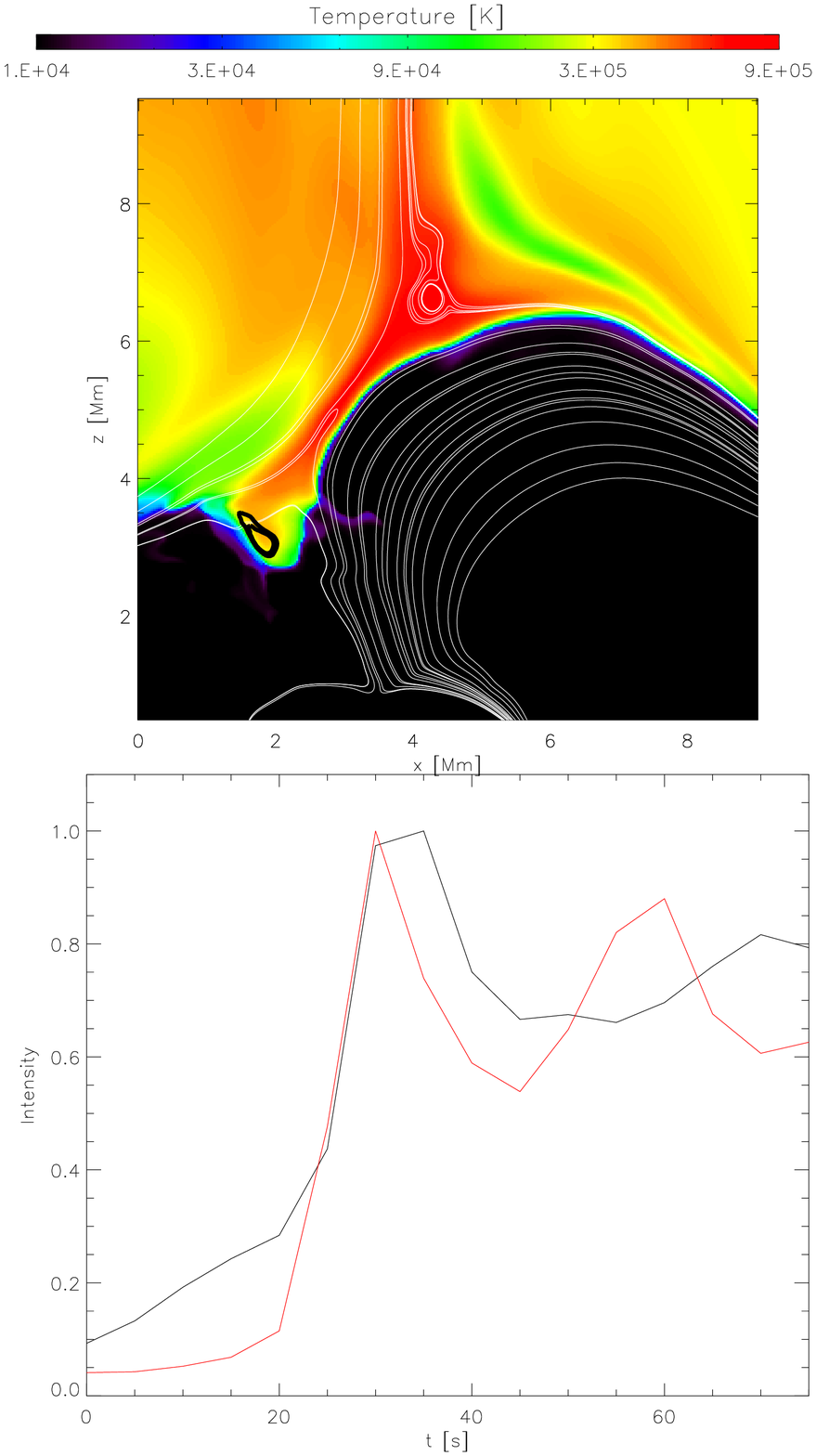}}\vspace{-0.4cm}
\caption{Top panel: Plasma temperature map of a snapshot of a 2D MHD
  model where a reconnection event occurs in the corona
  \citep{MartinezSykora12}. The snapshot selected is at the peak
  \ion{Fe}{8} emission, which is shown with the black contours around
  $[x,z] = [1.8,3.2]$~Mm. The magnetic field is shown with white
  lines.
  Bottom panel: Lightcurves of \ion{Fe}{8} (red) and \ion{Fe}{10}
  (black) emission averaged over an area corresponding to $3
  \times 3$ \hic\ pixels centered in the location where \ion{Fe}{8}
  emission is strongest ($x=1.8$~Mm).\label{fig:2DMHD}} 
\end{figure}

We find that rapid heating events on timescales comparable with the
\hic\ moss brightenings presented here are also observed in
MHD simulations.  We consider here 2D radiative MHD simulations with
thermal conduction along the magnetic field lines
\citep{MartinezSykora12}, performed using the Bifrost code
\citep{Gudiksen11}. This model shows a magnetic field configuration
with a reconnection X point in the corona (see
Figure~\ref{fig:2DMHD}).  The \ion{Fe}{8} and \fex\ emission,
synthesized under optically thin approximation and ionization
equilibrium, shows strong variations on timescales of roughly 15~s
(bottom panel of Figure~\ref{fig:2DMHD}).  
This rapid small scale variability is due to the heating timescales
and thermodynamic evolution along the field lines resulting from the
reconnection in the corona. We note that the modeled coronal
temperatures are significantly lower than the observed hot
loops. However, the models suggest that brightenings due to coronal
episodic heating can be observed with high spatial and temporal
resolution capabilities of instruments like \hic.

In conclusion we find that \hic\ observations of moss show evidence of
very rapid variability which we interpret as a signature of coronal
nanoflares. These observations showcase the diagnostic potential of
moss observations for coronal heating studies, when spatial and
temporal resolution are sufficiently high. 
Observations in several passbands sensitive to different temperatures,
for instance analogous to \aia\ passbands, at the level of spatial and
temporal resolution of \hic\ would provide significantly stronger
constraints to solve many of the open questions, e.g., the mechanism
of energy transport (conduction vs.\ beams).

\acknowledgments
We thank the referees for their useful comments which greatly helped
to improve the paper.  
PT was supported by contract SP02H1701R from Lockheed-Martin, and NASA
contract NNM07AB07C to SAO.   
BDP was supported through NASA grants NNX08BA99G, NNX08AH45G and
NNX11AN98G. 
\hinode\ is a Japanese mission developed and launched by ISAS/JAXA, with 
NAOJ, NASA and STFC (UK) as partners, and operated by these agencies
in co-operation with ESA and NSC (Norway).
SK was supported by RFBR project 11–02–01079-a, and Program22 of the RAS Presidium.

\end{document}